\documentclass[12pt,a4paper,dvips]{article}
\usepackage{epsfig}
\usepackage{a4p}
\usepackage{cite,mcite}
\usepackage{graphicx}
\usepackage{amssymb}
\usepackage{physics}
\usepackage{l3_title,ifthen,Lep}
\journalname{Phys. Lett. B}
\preprint{2001-096}
\date{December 17, 2001}
\Lep{2}
\newlength{\capindent}
\newlength{\capwidth}
\newlength{\figwidth}
\newcommand{\icaption}[2][!*!,!]{\hspace*{\capindent}%
  \begin{minipage}{\capwidth}
    \ifthenelse{\equal{#1}{!*!,!}}%
      {\caption{#2}}%
      {\caption[#1]{#2}}
  \end{minipage}}
\begin{document}
\begin{titlepage}

\title{\boldmath{Study of Multiphoton Final States and Tests of QED in $\epem$ collisions
       at $\sqrt{s}$ up to 209 GeV}}
\author{The L3 Collaboration}
%
\begin{abstract}
The process $\epem \ra n \gamma$ with $n\geq 2$ is studied
at centre-of-mass energies ranging from $\sqrt{s}=192$ to $209$ GeV. The 
data sample corresponds to a total integrated luminosity of 427 
pb$^{-1}$. The total and differential cross sections are found to be in 
agreement with the QED expectations. Using all the data collected with 
the L3 detector above the Z pole, limits on deviations from QED, 
excited electrons, contact interactions, extra space 
dimensions and excited spin-3/2 leptons are set.
\end{abstract}
\submitted
\end{titlepage}

\section*{Introduction}                                         

The process $\epem \ra \gamma\gamma$ receives its main contribution
from QED by means of the exchange of an electron via $t$-channel. The lowest 
order contribution to the cross section is:
\begin{eqnarray}
\left(\frac{d{\sigma}}{d{\Omega}}\right)_{QED} = \frac{\alpha^2}{s} 
\frac{ \left( 1 + \cos^2 \theta \right) } { \left( 1 -\cos^2 \theta \right)},
\end{eqnarray}

\noindent
where $\theta$ is the polar angle of the photon, $\alpha$ the 
electromagnetic coupling constant and $\rts$ the centre-of-mass energy of 
the collision.

The experimental signature of the final state is clean, allowing the 
analysis of event samples with negligible background.
The sensitivity of this process to deviations with respect to the QED 
predictions grows with $\rts$ and, in addition, non-QED contributions are 
small. Any deviation can be therefore interpreted as a sign of new 
physics. 

In this letter, results of the study of the process 
$\epem \ra n \gamma$ ($n\geq 2$) are presented. The analysis is
performed on the data collected by the L3 detector~\cite{l3rf} at 
centre-of-mass energies from 191.6 to 209.2 GeV, for a total integrated
luminosity of 427 pb$^{-1}$. The luminosities as a function of $\rts$ 
are detailed in Table~\ref{tab:lumeff}. L3 results at 
$\rts=91-189 \GeV$~\cite{gg91,gg130,gg172,gg189} are included in the 
interpretations. 
Similar studies at $\sqrt{s}$ up to $202\GeV$ were reported by other experiments~\cite{ggother}.

\section*{Event Selection}

  The event selection proceeds from photon candidates, defined as:

\begin{itemize}
\item A shower in the electromagnetic calorimeter with an energy above 5 GeV 
      having a profile consistent with that of a photon or an electron.

\item The number of hits in the vertex chamber within an azimuthal angle of
      $\pm 8^{\circ}$ around the path of the photon candidate must be less 
      than the 40\% of that expected for a charged particle. 
\end{itemize}

There must be at least two photon candidates with polar angles 
$\theta_{\gamma}$ between $16^{\circ}$ and $164^{\circ}$, for the shower 
to be fully contained in the electromagnetic calorimeter and to ensure a 
sufficient number of hits in the vertex chamber in order to reject electrons.
The angular separation between the two photons must be more than $15^{\circ}$. 
In addition, to reject $\epem \ra \nu \bar{\nu} \gamma \gamma$ and cosmic ray, 
events the sum of the energies of the photon candidates is required to be 
larger than $\sqrt{s}/2$. Events containing any track with momentum larger 
than 0.1 GeV pointing in a cone of $2.5^{\circ}$ around any additional 
calorimetric cluster are rejected. A scintillator signal in coincidence 
with the beam crossing time and associated to a photon is also required.

The background in the sample selected with these cuts, estimated from Monte 
Carlo simulations, is negligible. The efficiencies to detect at 
least two photons in the angular region 
$16^{\circ} < \theta_{\gamma} < 164^{\circ}$
are computed from a Monte Carlo generator~\cite{ggg} 
of $\epem \ra \gamma \gamma (\gamma)$ events of order $\alpha^3$, passed 
through the L3 simulation \cite{geant} and reconstruction programs. They are 
presented in Table~\ref{tab:lumeff}. Trigger inefficiencies, as 
estimated using Bhabha events, which have an independent trigger
for charged particles, are found to be negligible.

\section*{Analysis of the Sample}

\begin{figure}
\begin{center}
\begin{tabular}{cl}
\includegraphics[width=8.0truecm]{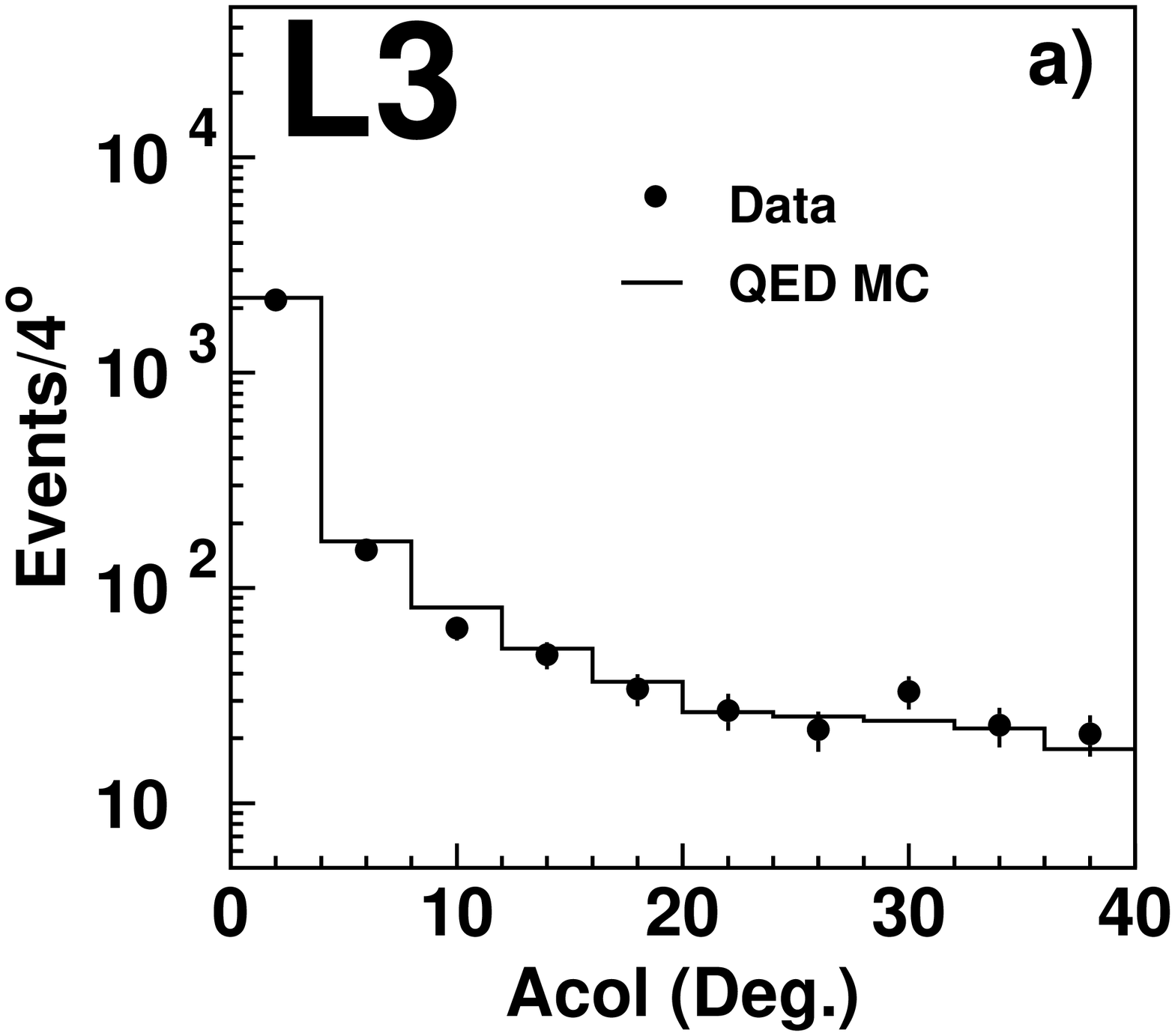}       &
\includegraphics[width=8.0truecm]{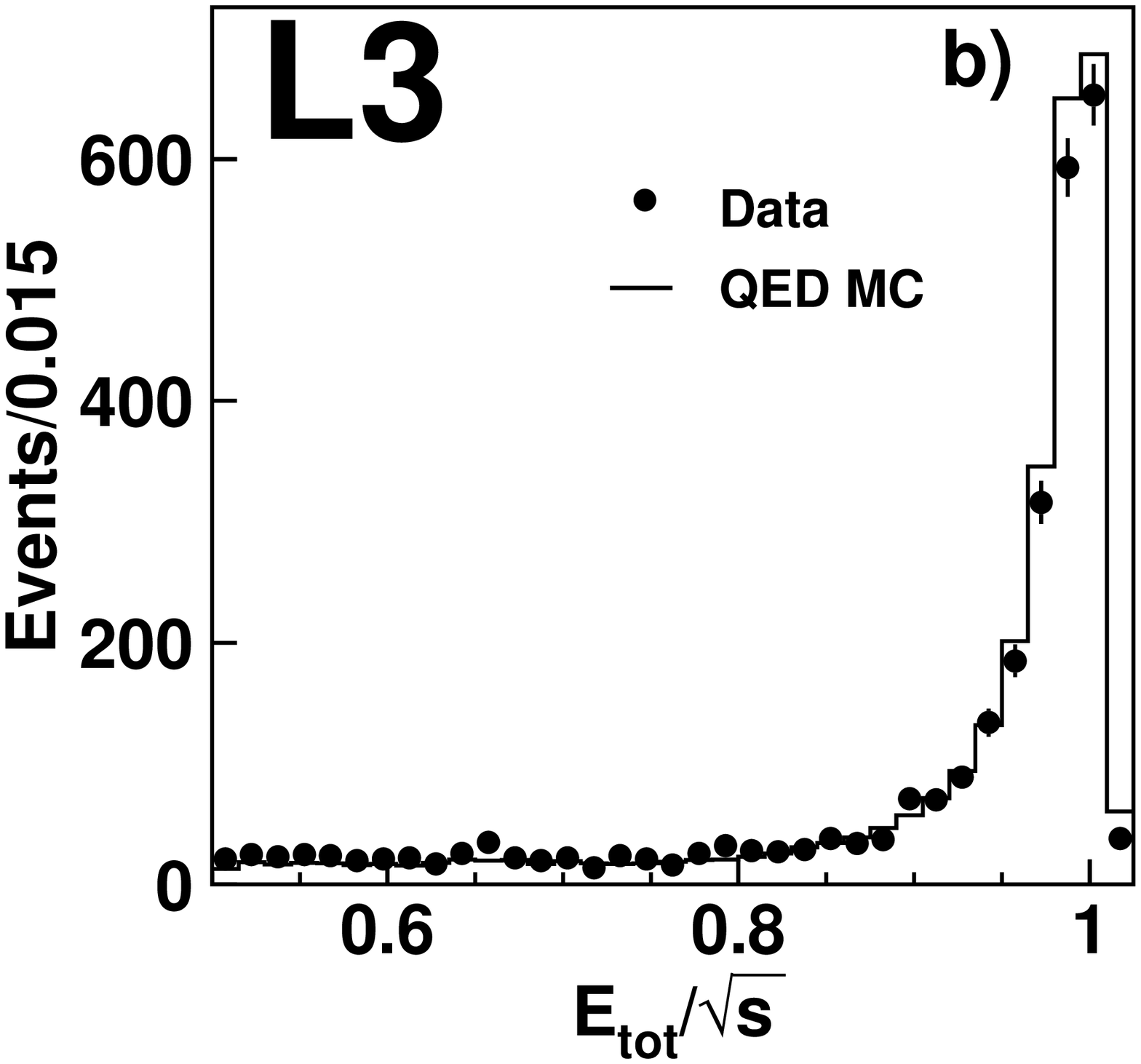}       \\
\includegraphics[width=8.0truecm]{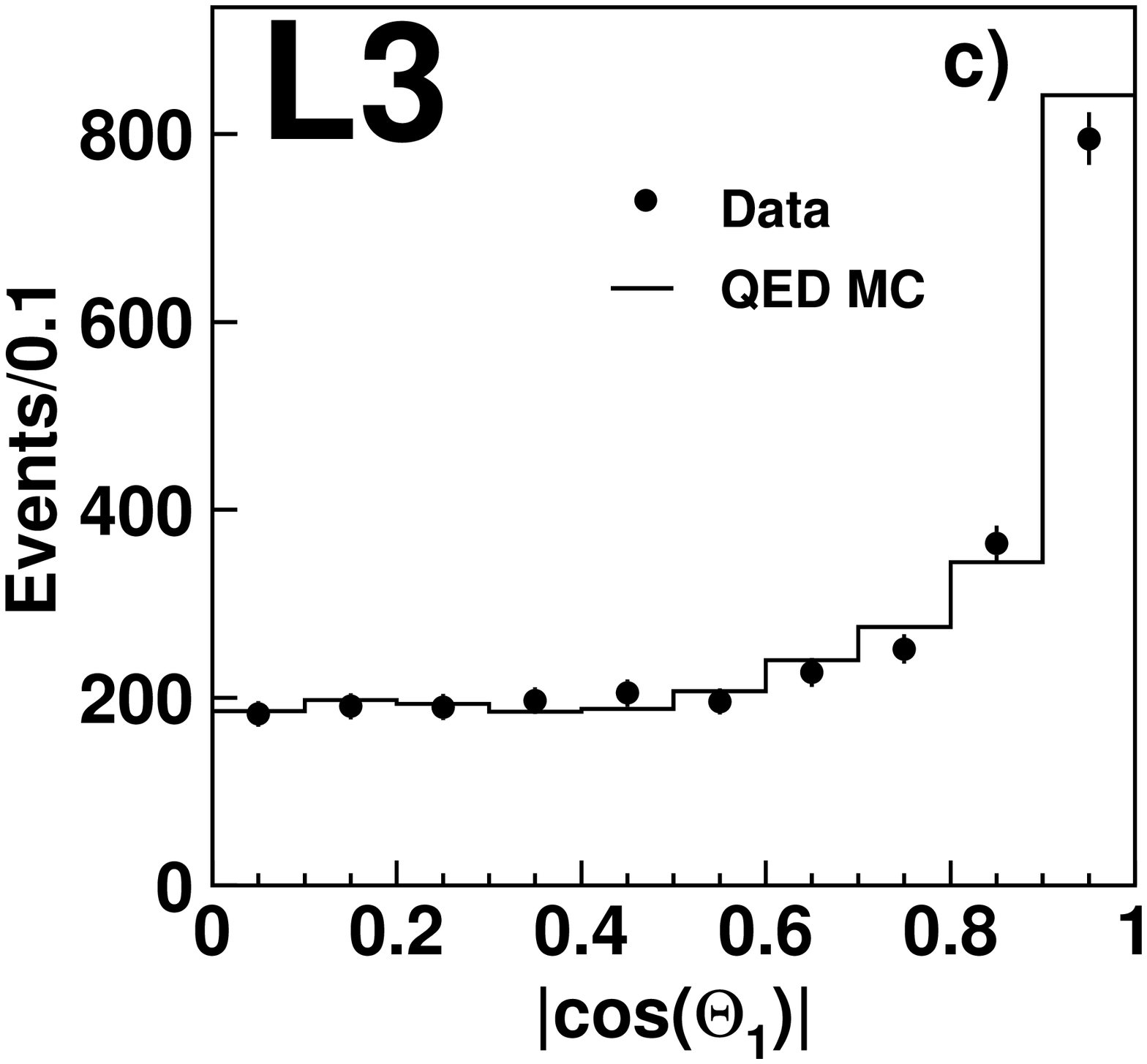}  &
\includegraphics[width=8.0truecm]{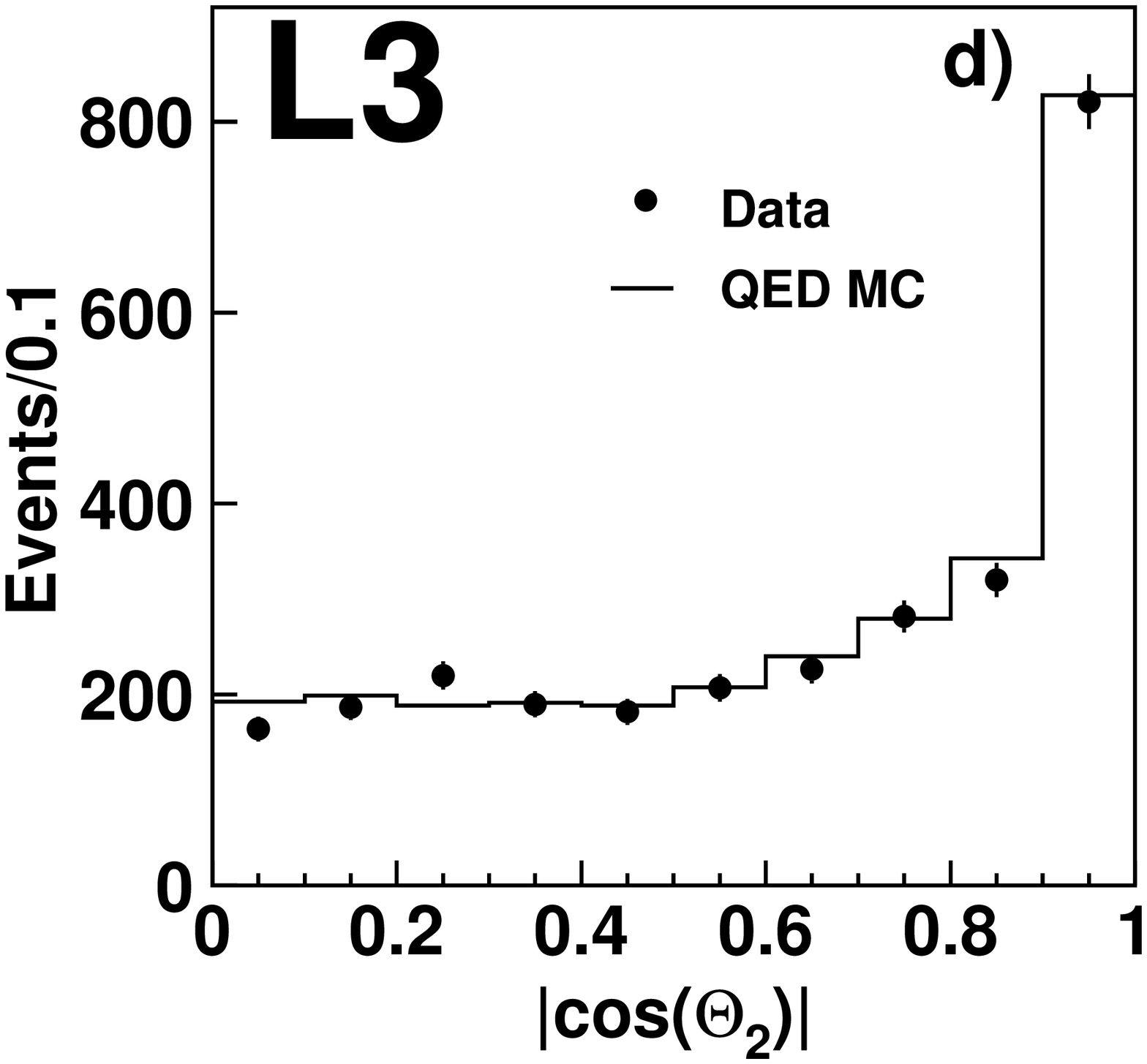}  \\
\end{tabular}
\end{center}
\icaption{\label{fig:distr}
    Distributions of a) the acollinearity angle between the two 
    most energetic photons, b) the total 
    energy normalized to the centre-of-mass energy and
    $\cos \theta$ for c) the most and d) the least energetic 
    photon. Points are data and the histogram is the Monte Carlo prediction.
    The data sample collected at $\rts=192-209 \GeV$ is presented.}
\end{figure}

After the selection criteria described above, events are classified 
according to the number of isolated photons in the angular range 
$16^{\circ} < \theta_{\gamma} < 164^{\circ}$. Table~\ref{tab:number_of_g}
lists the number of observed and expected events.
No events with four or more photons are observed
while 0.3 are expected~\cite{gg130}. One event with four photons was observed at 
$\sqrt{s}=130 \GeV$~\cite{gg130} and another one at 
$\sqrt{s}=183 \GeV$~\cite{gg189}. Integrating in 
the range $\sqrt{s}= 130-209 \GeV$, 0.7 of such events are expected.
The distributions of the acollinearity, the sum of the energies of the two most 
energetic photons and the polar angles 
of the most and least energetic photons are presented in 
Figure~\ref{fig:distr}. These distributions are obtained combining all data
at $\sqrt{s}=192-209 \GeV$. 

\begin{figure}
\begin{center}
\includegraphics[angle=90,width=19.5truecm]{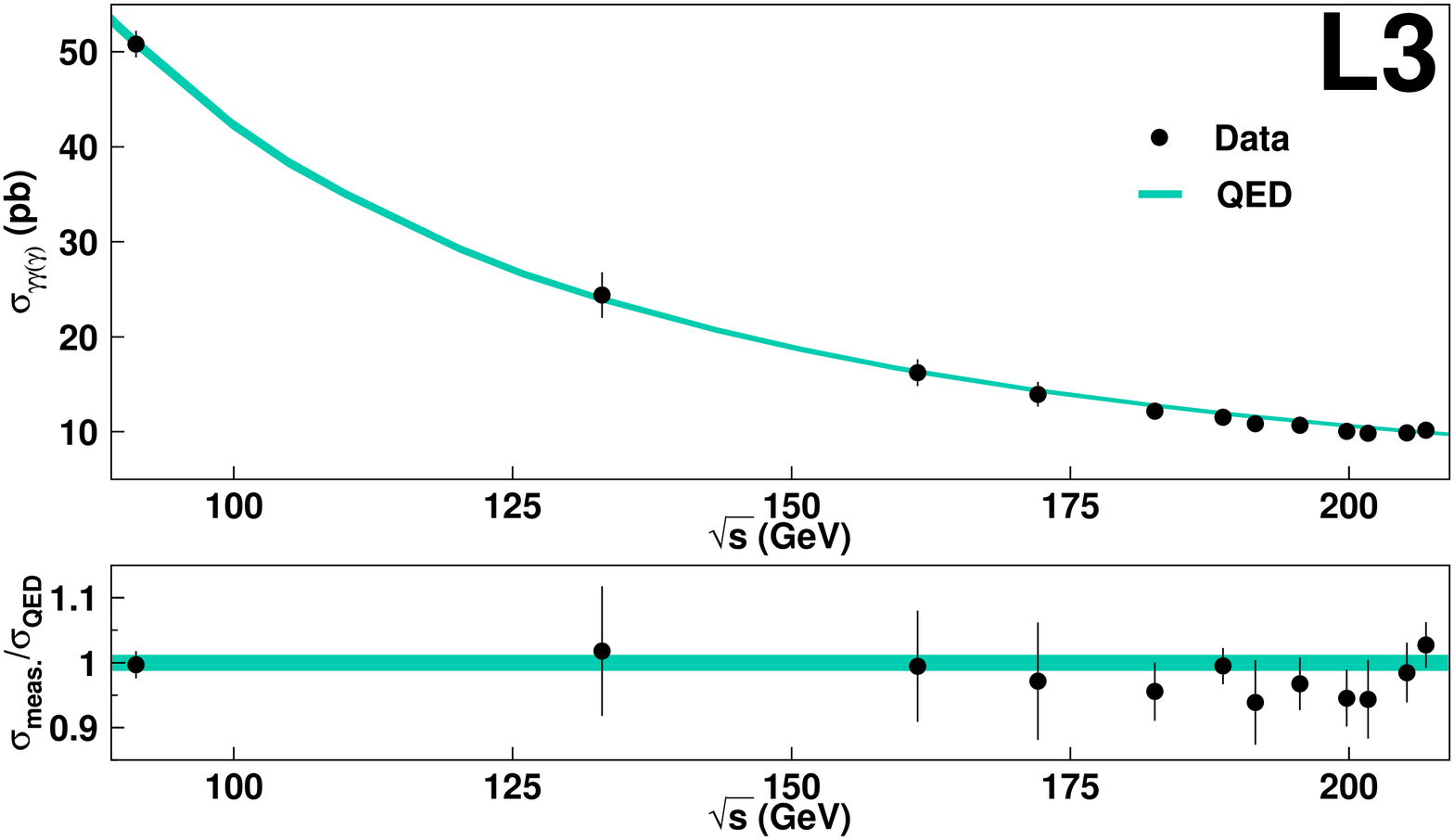}
\icaption{\label{fig:qedevol}
    Measured cross sections as a function of the centre-of-mass energy in 
    the angular region $16^{\circ} < \theta_{\gamma} < 164^{\circ}$,
    compared to QED predictions.
    The value at the Z pole is
    extrapolated to this angular range from the one given in 
    Reference~ {\protect \citen{gg91}}, resulting in a value of 
    $50.8 \pm 1.4$ pb. 
    The ratio between the measured and the expected cross sections is 
    also presented. The line width represents 
    the uncertainty in the QED prediction, estimated to be 1\%.}
\end{center}
\end{figure}

 The total cross sections are measured from the number of observed events. 
They are presented in Table~\ref{tab:xsecs} together with the QED 
expectations~\cite{ggg}. Good agreement is observed.
The uncertainty in the QED prediction, due to the missing contribution 
of higher order corrections, is estimated to be 1\%.
These measurements and the previously measured 
values~\cite{gg91,gg130,gg172,gg189} are presented in Figure~\ref{fig:qedevol} 
as a function of the centre-of-mass energy and compared to
the QED expectations. The
global $\chi^2$ of the data with respect to the theoretical prediction is
5.8 for 12 degrees of freedom, and the average ratio between the measured 
cross section, $\sigma_{measured}$, and the QED predicted cross section, 
$\sigma_{QED}$, is: $\sigma_{measured}/\sigma_{QED} = 
0.986 \pm 0.012 \pm 0.010$, where the first uncertainty is experimental 
and the second theoretical. 

The statistical uncertainty dominates the measurements. The main systematic 
source is the efficiency of the selection procedure. It
is evaluated varying the selection criteria and taking into account
the finite Monte Carlo statistics. The systematic effects due to the 
uncertainties in the measured luminosity and to the residual background 
are found to be negligible.

\begin{figure}
\begin{center}
\begin{tabular}{cc}
\includegraphics[width=7.0truecm]{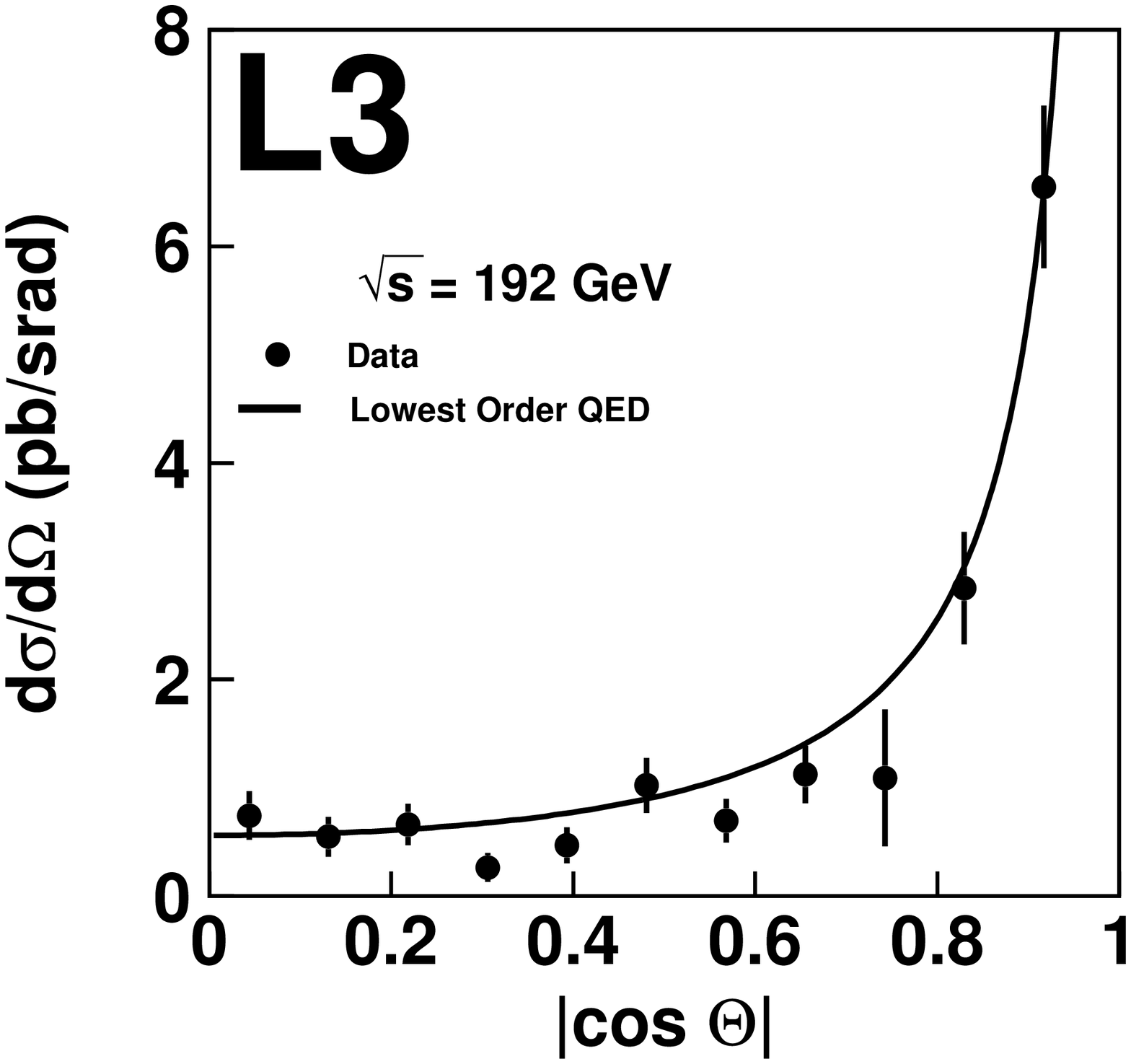} &
\includegraphics[width=7.0truecm]{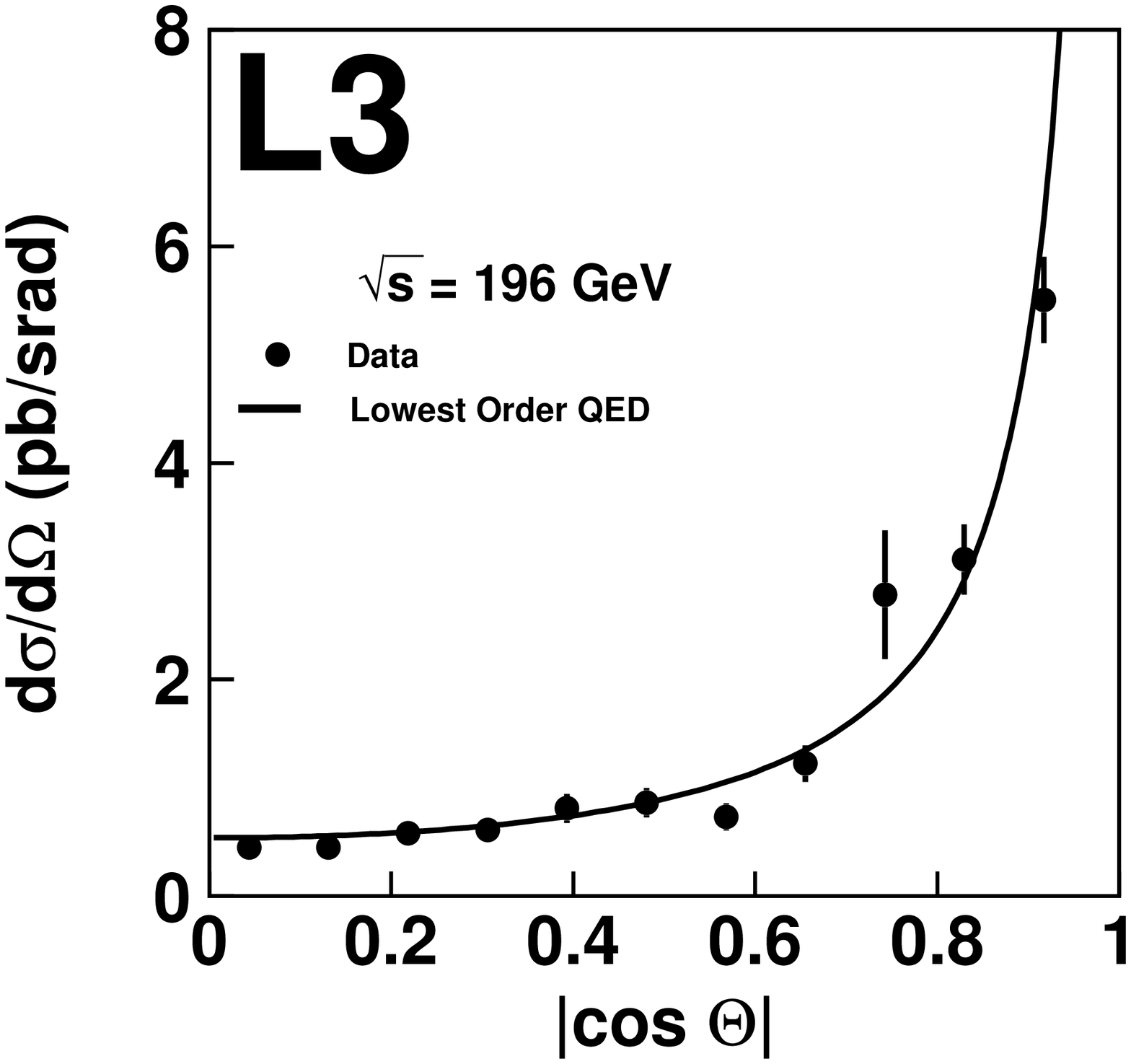} \\
\includegraphics[width=7.0truecm]{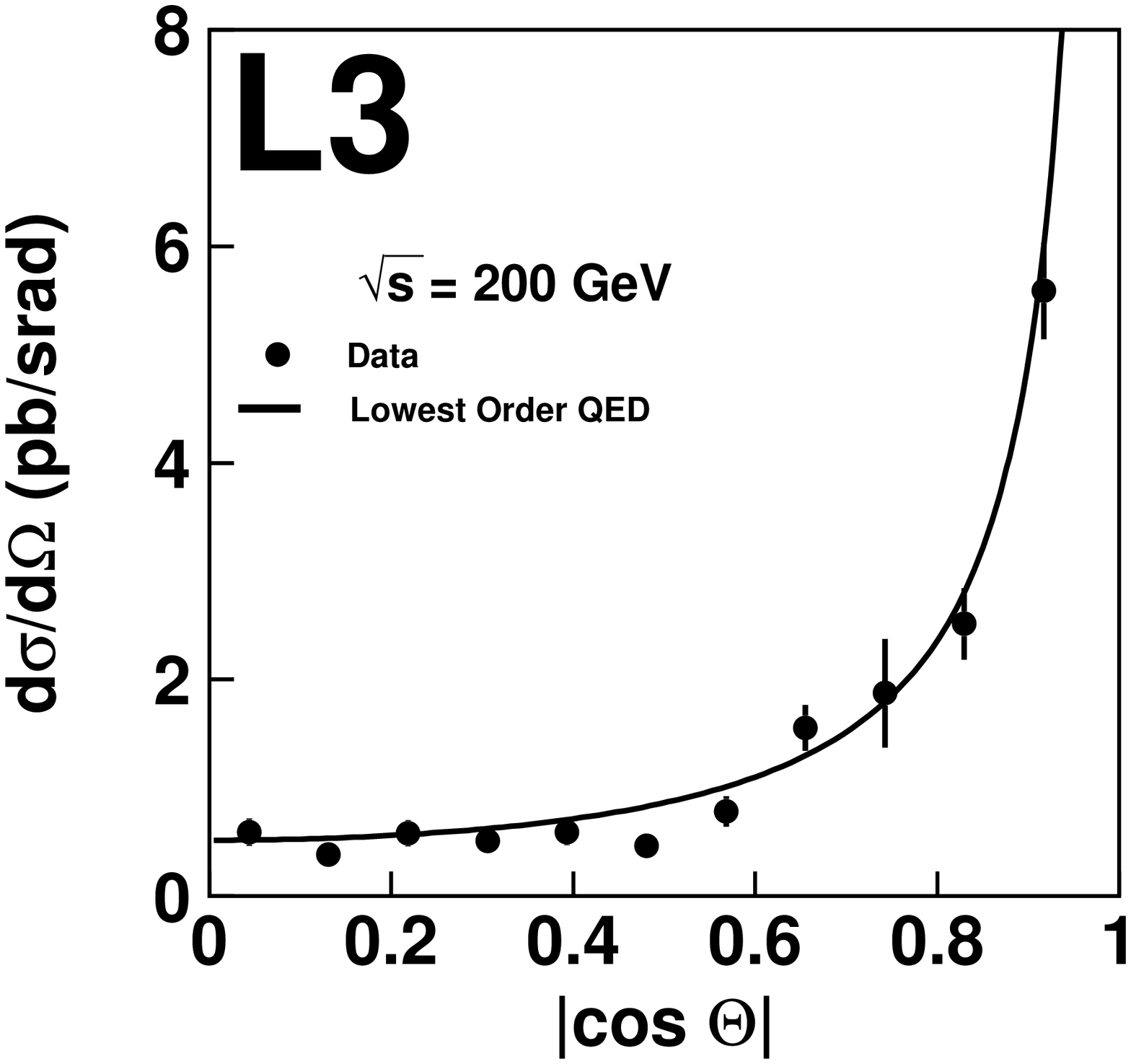} & 
\includegraphics[width=7.0truecm]{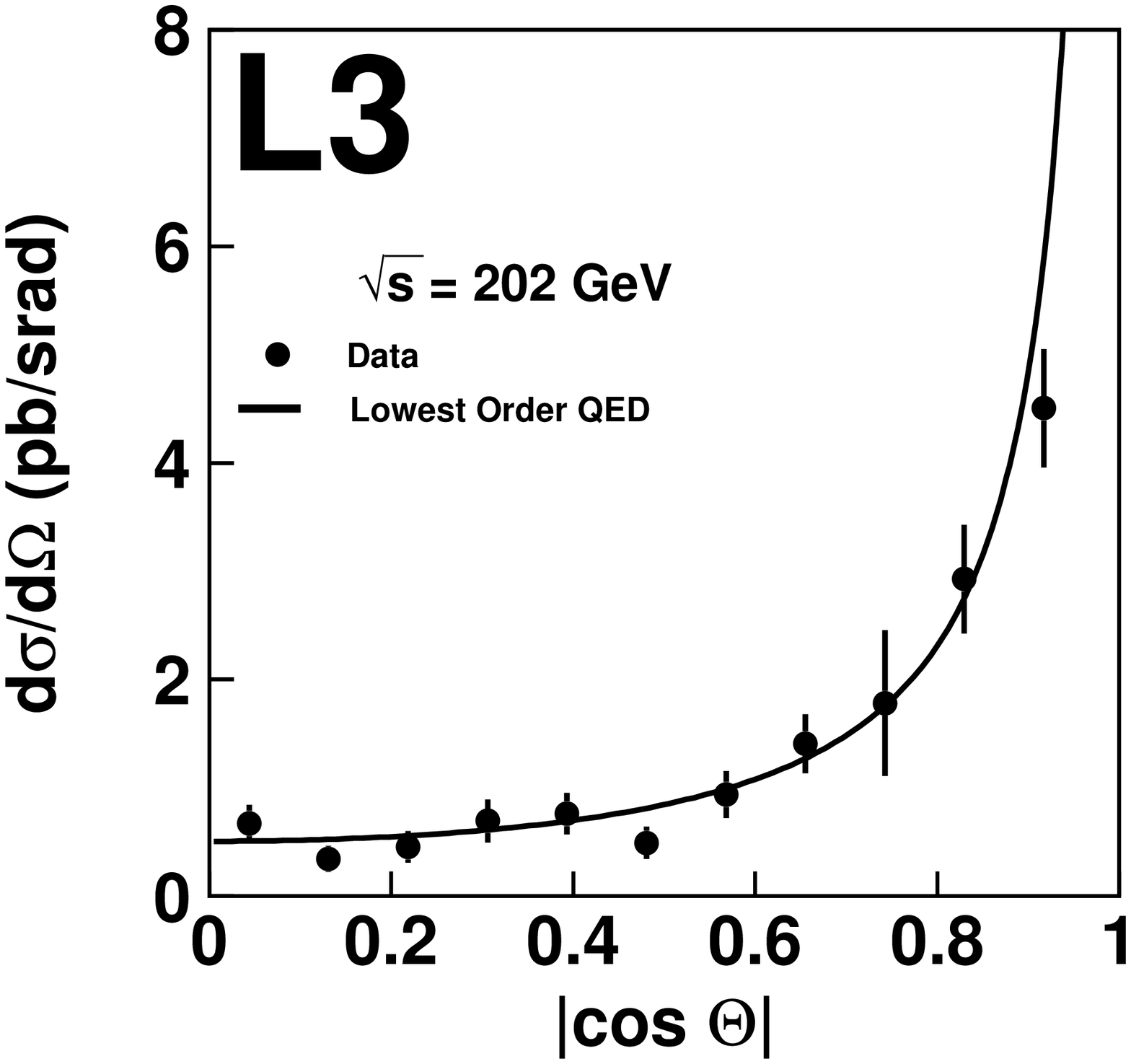} \\
\includegraphics[width=7.0truecm]{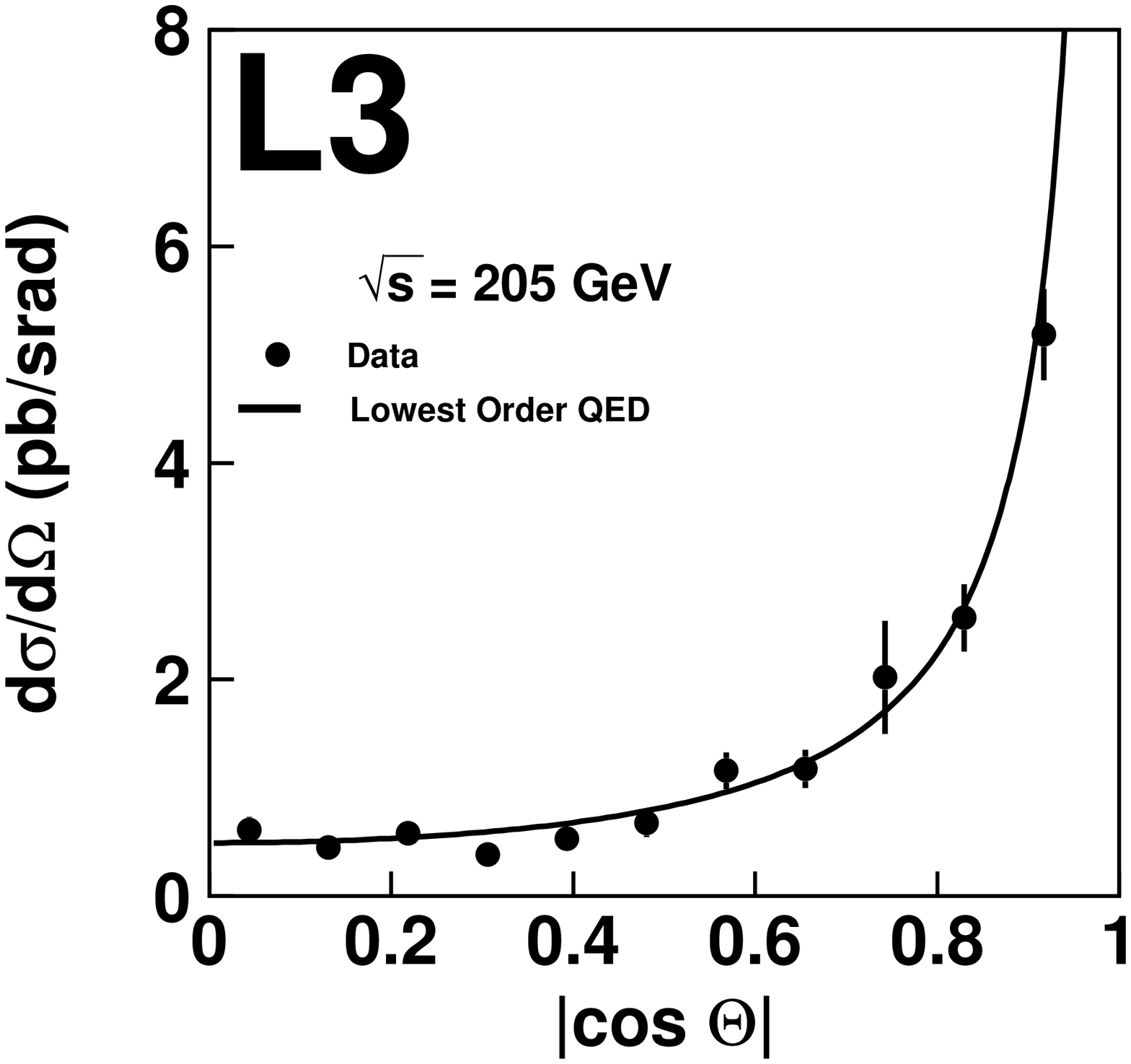} &
\includegraphics[width=7.0truecm]{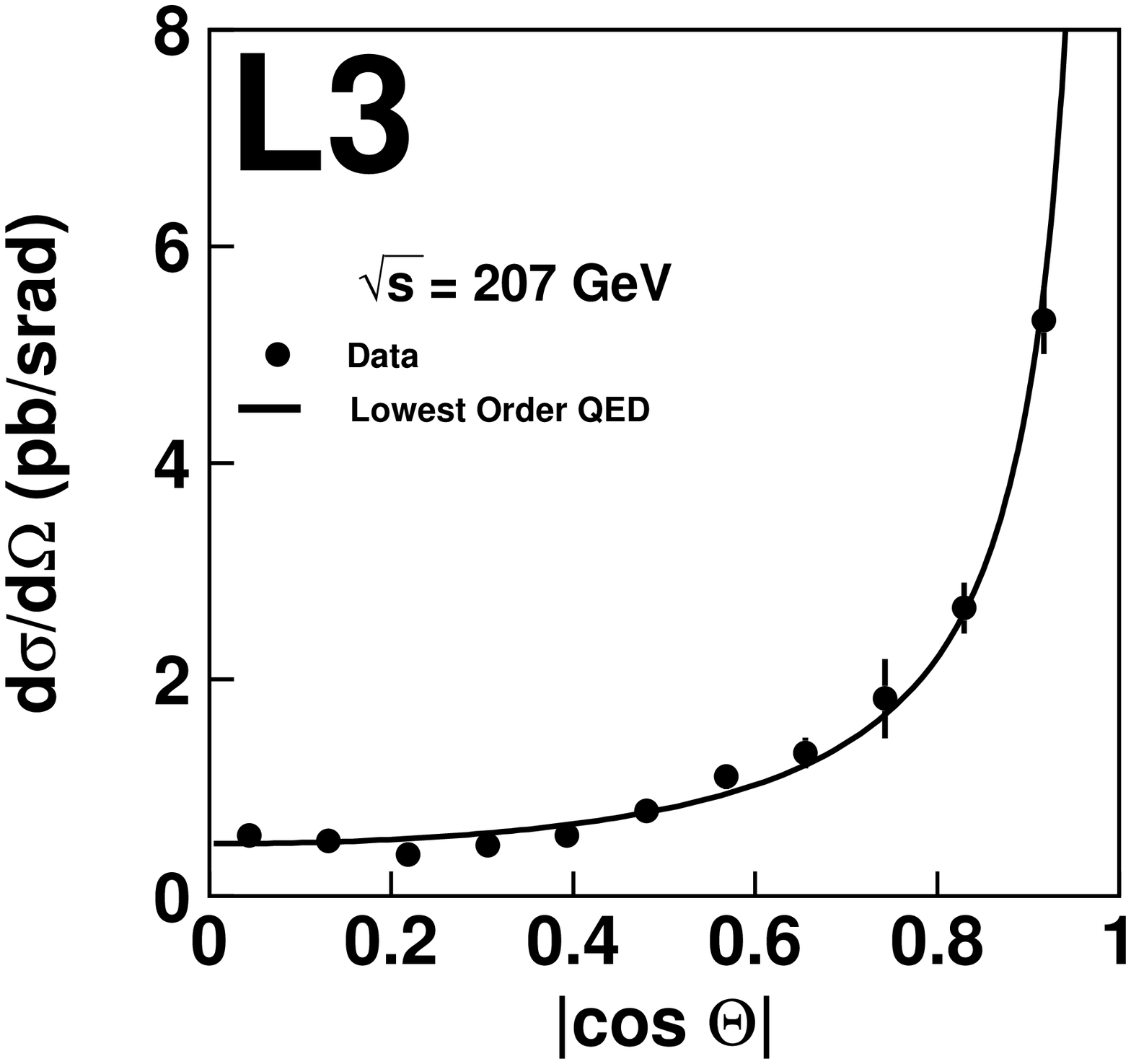} 
\end{tabular}
\end{center}
\icaption{\label{fig:costheta}
    Differential cross sections as a function of $\cos \theta$ for 
    different values of $\rts$. Points are data and the solid line 
    corresponds to the lowest order QED prediction.}
\end{figure}

The differential cross sections as a function of the polar angle 
are computed. The event polar angle, $\cos\theta$, is defined as
$\cos\theta= |\sin(\frac{\theta_1 - \theta_2}{2})/
\sin(\frac{\theta_1 + \theta_2}{2})|$,
where ${\theta}_1$ and ${\theta}_2$ are the polar angles of the two most 
energetic 
photons in the event. They are compared with the lowest order 
QED predictions for each $\sqrt{s}$ in Figure~\ref{fig:costheta}. A finer 
binning is presented in Table~\ref{tab:lep}. 
The table includes the bin-by-bin efficiencies and the multiplicative 
factors used to bring the cross section to the lowest order. 

  The agreement between data and expectations allows to constrain new physics
models. They are discussed in what follows.
 
\section*{Limits on Deviations from QED}

The possible deviations from QED are parametrised in terms of effective 
Lagrangians. Their effect on the observables is expressed as a 
multiplicative correction term to the QED differential cross-section. 
Depending on the type of parametrisation two general forms are 
considered~\cite{lambda}:
\begin{eqnarray}
\frac{d{\sigma}}{d{\Omega}}\,=\,
\left(\frac{d{\sigma}}{d{\Omega}}\right)_{QED}\,
\left(1+\frac{s^2}{\alpha} \frac{1}{{\Lambda}^4} {\rm sin}^2{\theta} \right)
\label{eqn1}
\end{eqnarray}

\noindent and
\begin{eqnarray}
\frac{d{\sigma}}{d{\Omega}}\,=\,
\left(\frac{d{\sigma}}{d{\Omega}}\right)_{QED}\,
\left(1+\frac{s^3}{32{\pi}{\alpha}^2} \frac{1}{{\Lambda}^{'6}}
\frac{{\rm sin}^2{\theta}}{1+{\rm cos}^2{\theta}} \right),
\end{eqnarray}

\noindent
which depend on the centre-of-mass energy, the polar angle ${\theta}$ and the 
scale parameters $\Lambda$ or $\Lambda^{'}$. A simple and convenient way 
of parametrising the deviations from QED is the introduction of the cut-off 
parameters ${\Lambda}_{\pm}$~\cite{cutoff}. The differential cross-section 
in this case is obtained from Equation~\ref{eqn1} replacing 
${\Lambda}^4$ by ${\pm}(2/{\alpha}){\Lambda}^4_{\pm}$. The
effects of deviations of this type on the differential cross section are 
presented in Figure~\ref{fig:qed_deviations}.

 Combining the present results with those obtained in our previous 
analyses~\cite{gg130,gg172,gg189}, the estimated parameters are:
\begin{eqnarray}
 \frac{1}{{\Lambda}^4}\, &=& \left(-0.01^{+\, 0.03}
   _{-\, 0.02} \right)\,\times 10^{-11}\,\,\,\,\,{\rm GeV}^{-4},  \nonumber \\
 \frac{1}{{{\Lambda}^{'}}^6}\, &=& \left(-0.03^{+\, 0.06}
     _{-\, 0.04} \right)\,\times 10^{-16}\,\,\,\,\,{\rm GeV}^{-6}.  \nonumber
\end{eqnarray}

  Normalising the corresponding probability density function over the 
physically allowed range of the parameters, the following limits at the 
95\% confidence level (CL) are obtained:

\begin{center}
\begin{tabular}{lcr}
$\Lambda     $  & $>$ & 1.6 ~~TeV, \\
$\Lambda_+   $  & $>$ & 0.4 ~~TeV, \\
$\Lambda_-   $  & $>$ & 0.3 ~~TeV, \\
$\Lambda^{'} $  & $>$ & 0.8 ~~TeV. \\
\end{tabular}
\end{center}

\section*{Search for Excited Electrons}

  Another way to study possible deviations from QED is to postulate the 
existence of an excited electron $\mathrm{e}^*$ of mass $m_{\mathrm e^*}$, 
which couples to 
the electron and the photon via chiral magnetic interactions. 
This coupling is described by the phenomenological 
Lagrangian~\cite{estr2}:
\begin{eqnarray}
{\cal{L}}\,=\,\frac{e}{2{\Lambda}_{\mathrm e^*}}\,
\overline{\Psi}_{\mathrm e^*} {\sigma}^{\mu\nu} (1 \pm {\gamma}^5)
         {\Psi}_{\mathrm e} F_{\mu\nu}\,+\, h.c.
\end{eqnarray}

The parameter ${\Lambda}_{\mathrm e^*}$ is related to the effective scale 
of the interaction. The effect on the differential cross section due to the
presence of an excited electron with ${\Lambda}_{\mathrm e^*}=m_{\mathrm e^*}$ is depicted
in Figure~\ref{fig:qed_deviations}. From a fit to the data, we obtain:
\begin{eqnarray}
 \frac{1}{\Lambda_{\mathrm e^*}^4}\, &=& \left(-0.09^{+\, 0.20}
     _{-\, 0.17} \right)\,\times 10^{-9}\,\,\,\,\,{\rm GeV}^{-4}.  \nonumber
\end{eqnarray}

Fixing the interaction scale ${\Lambda}_{\mathrm e^*}$ to $m_{\mathrm e^*}$, 
we derive a $95\%$ CL lower limit of:
\begin{eqnarray}
 m_{\mathrm e^*}                &>& 0.31\,\,\,\,\,{\rm TeV}. \nonumber
\end{eqnarray}

 No excited electron mass limit with a purely magnetic 
interaction~\cite{estr1} is given, since the limits derived from $g_e -2$ 
measurements already exclude~\cite{g-2} the scales accessible at LEP. 

\section*{Low Scale Gravity Effects}

The differential cross section for photon pair production in $\epem$ 
collisions is modified in the presence of Low Scale Gravity and extra space 
dimensions~\cite{gravex1, gravex2}. From Reference~\citen{gravex2} it 
follows:
\begin{eqnarray}
 \frac{d \sigma}{d\Omega}
= \left(\frac{d{\sigma}}{d{\Omega}}\right)_{QED}\,
\left( 1 - \frac{\lambda s^2} {2 \pi \alpha M_S^4} \left( 1 - \cos^2 \theta \right) 
         + \frac{\lambda^2 s^4} {16 \pi^2 \alpha^2 M_S^8} \left( 1 - \cos^2 \theta \right)^2 \right).
\label{formula:gggravity}
\end{eqnarray}

 The deviations are weighted by a factor $\lambda$ which
absorbs the full dependence on the details of the theory. The parameter 
$\lambda = \pm 1$ is chosen to allow for the different signs of the 
interference. The pure gravitational part in the third term never exceeds 
1\% of the second term, the interference one, and is thus neglected. From a 
comparison of Equations~ \ref{eqn1} and ~\ref{formula:gggravity} it 
follows:
\begin{eqnarray}
- \frac{\lambda}{M_S^4} = \pm \frac{\pi \alpha}{{\Lambda}^{4}_{\pm}}. \nonumber
\end{eqnarray}

 The modified differential cross section is shown in 
Figure~\ref{fig:qed_deviations}. Lower limits at 95\% CL on the value of the 
scale $M_S$, derived from 
the limits on $\Lambda_\pm$, are:

\begin{center}
\begin{tabular}{lcr}
 $M_{S}(\lambda = +1)$  & $>$ & 0.84 TeV, \\
 $M_{S}(\lambda = -1)$  & $>$ & 0.99 TeV. \\
\end{tabular}
\end{center}

\section*{Search for Excited Spin-3/2 Leptons}

\begin{figure}
\begin{center}
\includegraphics[width=15.7truecm]{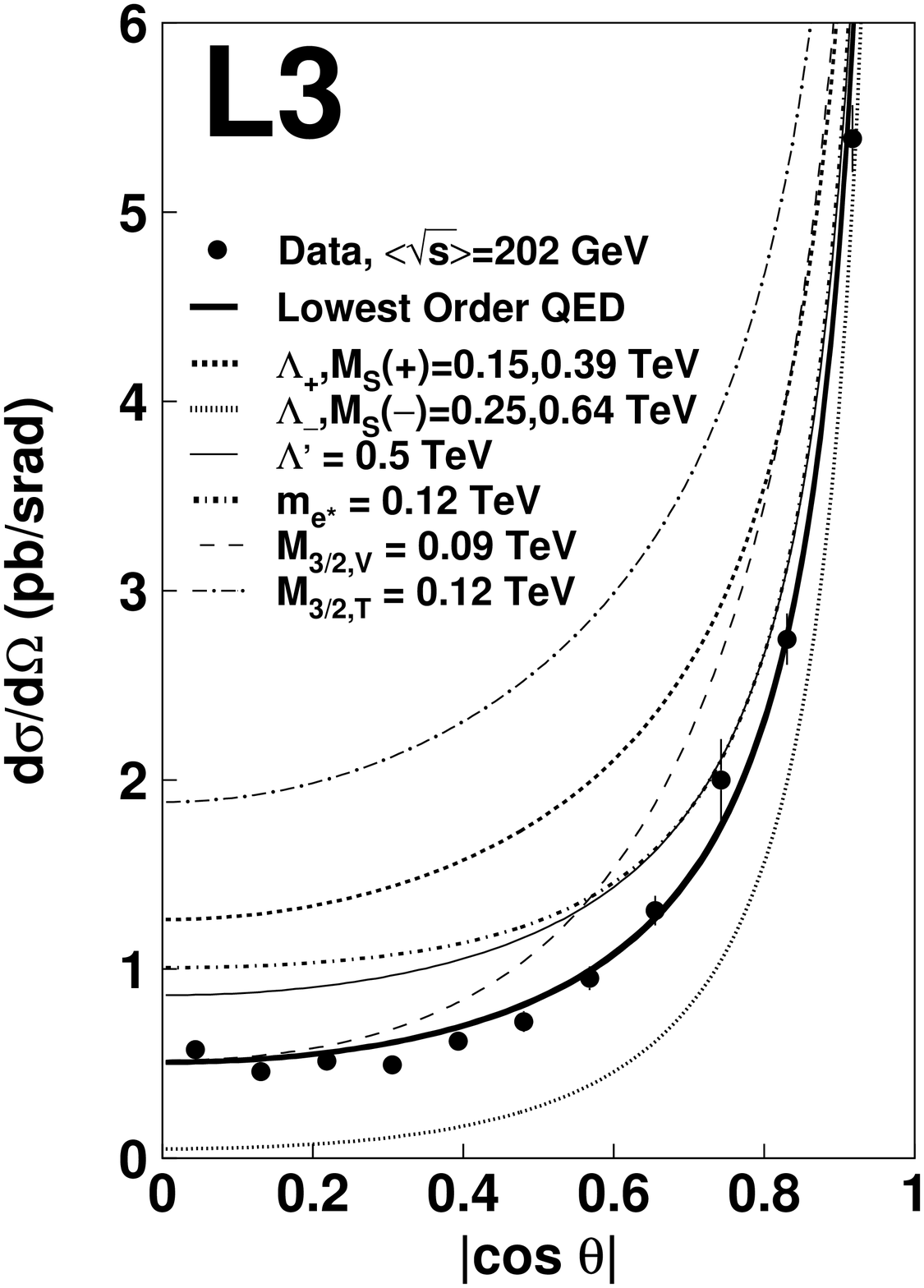} 
\end{center}
\begin{center}
\vspace*{-1.7truecm}
\icaption{\label{fig:qed_deviations}
    Differential cross sections as a function of $\cos \theta$. Points are
    data from $\sqrt{s}=192$ to $208 \GeV$, corresponding to 
    a luminosity weighted average of $<\sqrt{s}>=202 \GeV$. 
    Lines show the different predictions for the models discussed in the 
    text at a centre-of-mass energy of $<\sqrt{s}>=202 \GeV$. The width of 
    the lowest order QED prediction takes into account the theoretical 
    uncertainty, estimated to be 1\%. The $\chi^2$ with respect to the QED
    prediction is 1.6 per degree of freedom.}
\end{center}
\end{figure}

Supersymmetry and composite models~\cite{s32comp} predict the existence
of spin-3/2 particles, and $\ee\ra \gamma\gamma$ production is a 
suitable
process to search for their effect. Field theories for spin-3/2 particles
are known to be non-renormalizable, but two effective interaction 
Lagrangians~\cite{spin32}, with vector or tensor interactions, can be used to 
describe this contribution:
\begin{eqnarray}
{\cal L}^{(1)}_{int} & = & {e \over M_{3/2,V}} {\bar{\Psi}}^*_{\mu} 
{\gamma}_{\nu} 
(c_L\psi_L + c_R\psi_R) F^{\mu\nu} \mbox{ ,} \nonumber \\
{\cal L}^{(2)}_{int} & = & {e \over M_{3/2,T}^2} {\bar{\Psi}}^*_{\mu} 
{\sigma}_{\alpha\beta} (c_L\psi_L + c_R\psi_R)\partial^{\mu} 
F^{\alpha\beta} \mbox{ ,} 
\end{eqnarray}

\noindent
where $\Psi_{\mu}$ refers to the spin-3/2 lepton, $\psi_{L}$ and 
$\psi_{R}$ are the left and right handed electron fields, respectively, 
$c_L$ and $c_R$ are the corresponding coupling strengths, and $F^{\mu\nu}$ 
the electromagnetic field tensor. The parameters $M_{3/2,i}$ ($i=V,T$)
are the masses of the excited lepton for each hypothesis, and are also 
identified with the scale of new physics. The presence of such lepton 
modifies the differential cross section of 
the $\epem \ra \gamma  \gamma$ process as presented in 
Figure~\ref{fig:qed_deviations}.

\begin{figure}
\begin{center}
\includegraphics[width=15.0truecm]{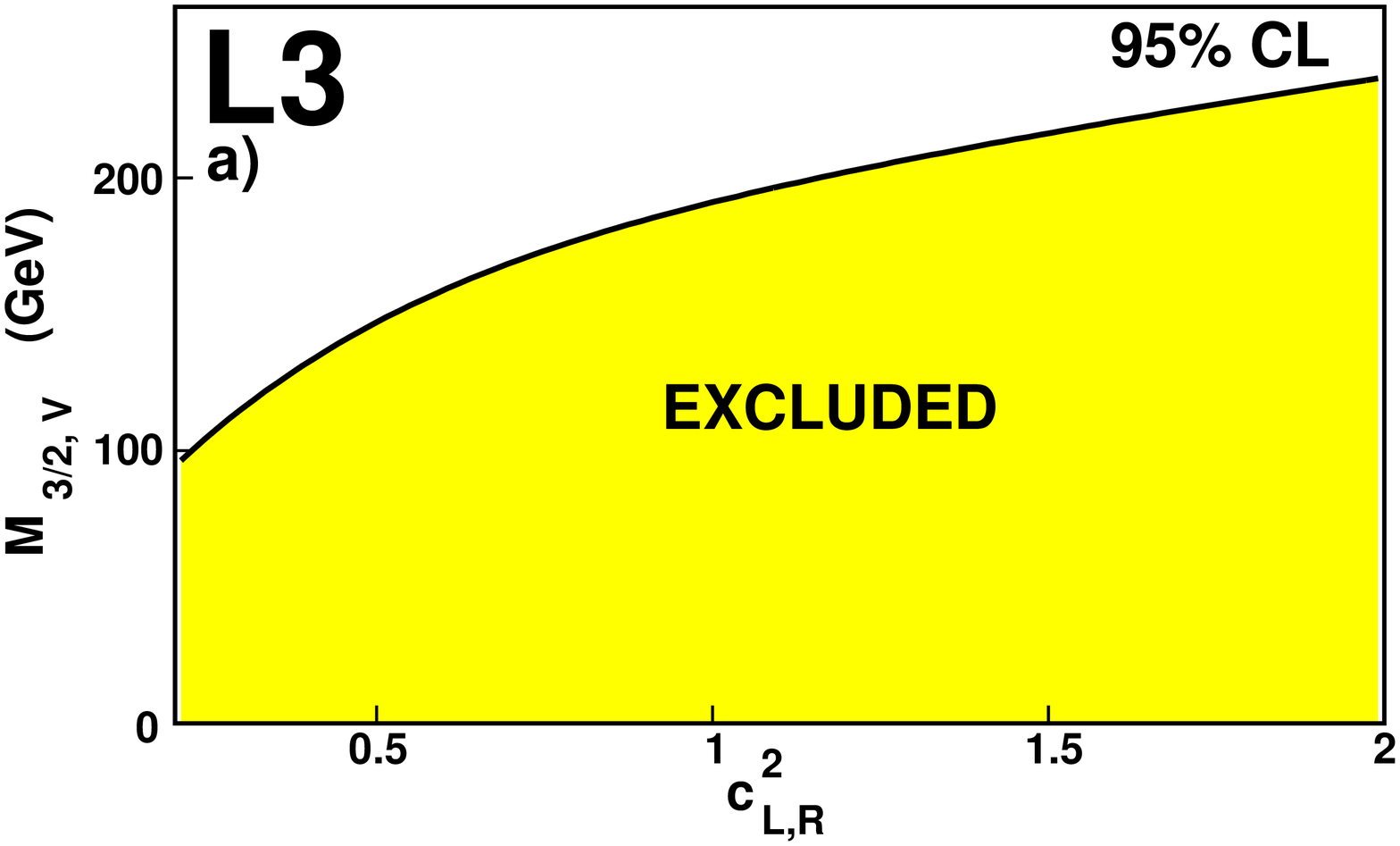} 
\includegraphics[width=15.0truecm]{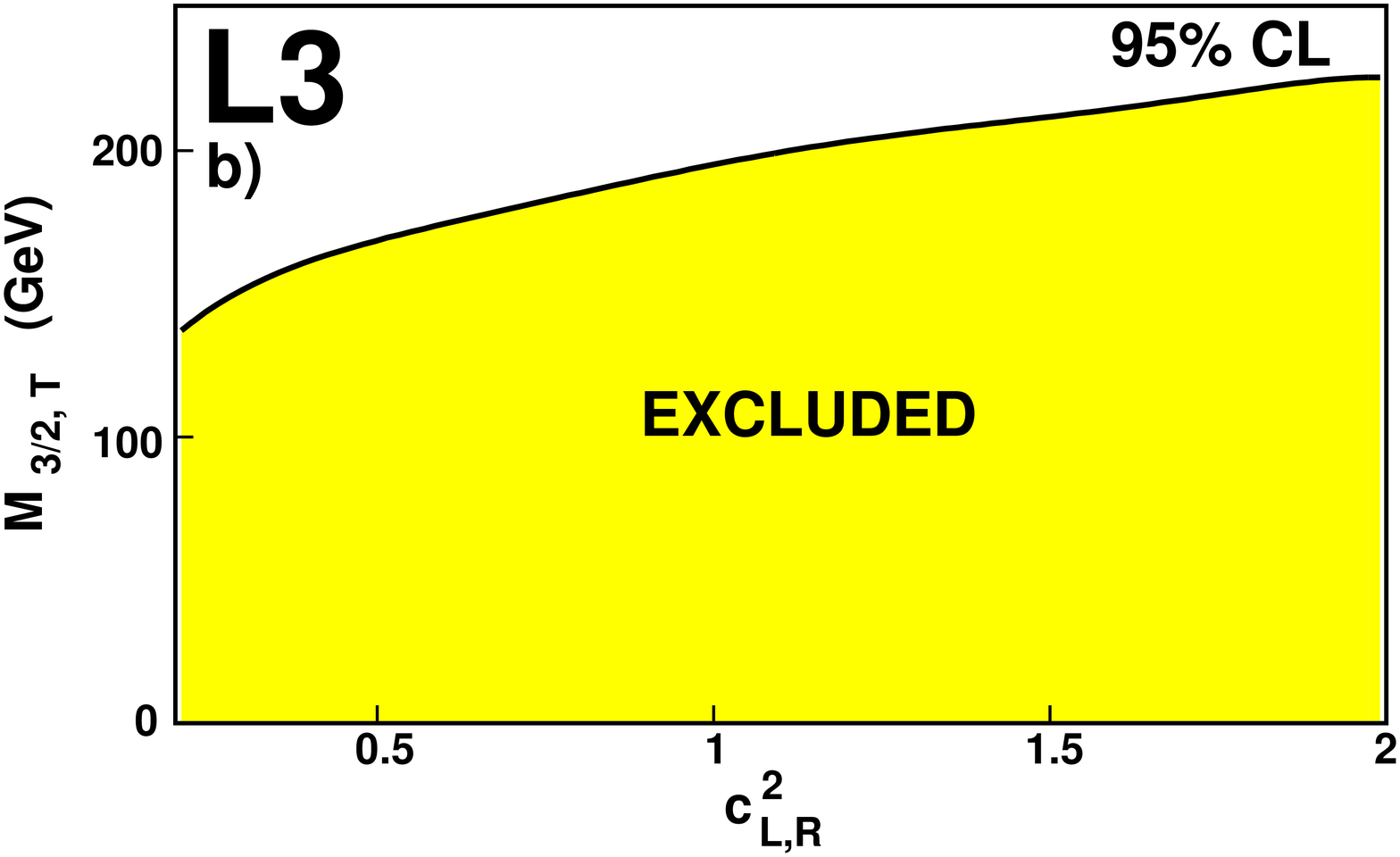}
\icaption{\label{fig:mvscl32}
    Excluded regions at 95\% CL in the plane a) $(M_{3/2, V}^2, c_L^2)$ 
    for the vector coupling case and b) $(M^2_{3/2, T}, c_L^2)$ 
    for the tensor coupling hypothesis in the search for excited spin-3/2 
    leptons. 
    The result is independent of the interchange between $c_L$ and $c_R$
    {\protect \cite{spin32}}.}
\end{center}
\end{figure}

 A search for excited spin-3/2 leptons is performed using all data collected 
with L3 above the Z pole under the assumption $c_{R}=0$. Deviations from QED
are invariant under the interchange between $c_L$ and $c_R$~\cite{spin32}.
Figure~\ref{fig:mvscl32} presents the 95\% CL excluded regions in 
the $(c_{L}^2, M_{3/2,i})$ planes. The 95\% CL limits obtained 
for $c_{L}^2 =1$ are:

\begin{center}
\begin{tabular}{lcr}
 $M_{3/2,V}$  & $>$ & 0.19 TeV, \\
 $M_{3/2,T}$  & $>$ & 0.20 TeV. \\
\end{tabular}
\end{center}

\newpage
\bibliographystyle{l3style}

\newpage
\typeout{   }     
\typeout{Using author list for paper 249 -- ? }
\typeout{$Modified: Jul 31 2001 by smele $}
\typeout{!!!!  This should only be used with document option a4p!!!!}
\typeout{   }
%
%
%
%
%
%

\newcount\tutecount  \tutecount=0
\def\tutenum#1{\global\advance\tutecount by 1 \xdef#1{\the\tutecount}}
\def\tute#1{$^{#1}$}
\tutenum\aachen            
\tutenum\nikhef            
\tutenum\mich              
\tutenum\lapp              
\tutenum\basel             
\tutenum\lsu               
\tutenum\beijing           
\tutenum\berlin            
\tutenum\bologna           
\tutenum\tata              
\tutenum\ne                
\tutenum\bucharest         
\tutenum\budapest          
\tutenum\mit               
\tutenum\panjab            
\tutenum\debrecen          
\tutenum\florence          
\tutenum\cern              
\tutenum\wl                
\tutenum\geneva            
\tutenum\hefei             
\tutenum\lausanne          
\tutenum\lyon              
\tutenum\madrid            
\tutenum\florida           
\tutenum\milan             
\tutenum\moscow            
\tutenum\naples            
\tutenum\cyprus            
\tutenum\nymegen           
\tutenum\caltech           
\tutenum\perugia           
\tutenum\peters            
\tutenum\cmu               
\tutenum\potenza           
\tutenum\prince            
\tutenum\riverside         
\tutenum\rome              
\tutenum\salerno           
\tutenum\ucsd              
\tutenum\sofia             
\tutenum\korea             
\tutenum\purdue            
\tutenum\psinst            
\tutenum\zeuthen           
\tutenum\eth               
\tutenum\hamburg           
\tutenum\taiwan            
\tutenum\tsinghua          

{
\parskip=0pt
\noindent
{\bf The L3 Collaboration:}
\ifx\selectfont\undefined
 \baselineskip=10.8pt
 \baselineskip\baselinestretch\baselineskip
 \normalbaselineskip\baselineskip
 \ixpt
\else
 \fontsize{9}{10.8pt}\selectfont
\fi
\medskip
\tolerance=10000
\hbadness=5000
\raggedright
\hsize=162truemm\hoffset=0mm
\def\r{\rlap,}
\noindent

P.Achard\r\tute\geneva\ 
O.Adriani\r\tute{\florence}\ 
M.Aguilar-Benitez\r\tute\madrid\ 
J.Alcaraz\r\tute{\madrid,\cern}\ 
G.Alemanni\r\tute\lausanne\
J.Allaby\r\tute\cern\
A.Aloisio\r\tute\naples\ 
M.G.Alviggi\r\tute\naples\
H.Anderhub\r\tute\eth\ 
V.P.Andreev\r\tute{\lsu,\peters}\
% ...
\newpage
\begin{list}{A}{\itemsep=0pt plus 0pt minus 0pt\parsep=0pt plus 0pt minus 0pt
                \topsep=0pt plus 0pt minus 0pt}
\item[\aachen]
 I. Physikalisches Institut, RWTH, D-52056 Aachen, FRG$^{\S}$\\
 III. Physikalisches Institut, RWTH, D-52056 Aachen, FRG$^{\S}$
\item[\nikhef] National Institute for High Energy Physics, NIKHEF, 
     and University of Amsterdam, NL-1009 DB Amsterdam, The Netherlands
\item[\mich] University of Michigan, Ann Arbor, MI 48109, USA
\item[\lapp] Laboratoire d'Annecy-le-Vieux de Physique des Particules, 
     LAPP,IN2P3-CNRS, BP 110, F-74941 Annecy-le-Vieux CEDEX, France
\item[\basel] Institute of Physics, University of Basel, CH-4056 Basel,
     Switzerland
\item[\lsu] Louisiana State University, Baton Rouge, LA 70803, USA
\item[\beijing] Institute of High Energy Physics, IHEP, 
  100039 Beijing, China$^{\triangle}$ 
\item[\berlin] Humboldt University, D-10099 Berlin, FRG$^{\S}$
\item[\bologna] University of Bologna and INFN-Sezione di Bologna, 
     I-40126 Bologna, Italy
\item[\tata] Tata Institute of Fundamental Research, Mumbai (Bombay) 400 005, India
\item[\ne] Northeastern University, Boston, MA 02115, USA
\item[\bucharest] Institute of Atomic Physics and University of Bucharest,
     R-76900 Bucharest, Romania
\item[\budapest] Central Research Institute for Physics of the 
     Hungarian Academy of Sciences, H-1525 Budapest 114, Hungary$^{\ddag}$
\item[\mit] Massachusetts Institute of Technology, Cambridge, MA 02139, USA
\item[\panjab] Panjab University, Chandigarh 160 014, India.
\item[\debrecen] KLTE-ATOMKI, H-4010 Debrecen, Hungary$^\P$
\item[\florence] INFN Sezione di Firenze and University of Florence, 
     I-50125 Florence, Italy
\item[\cern] European Laboratory for Particle Physics, CERN, 
     CH-1211 Geneva 23, Switzerland
\item[\wl] World Laboratory, FBLJA  Project, CH-1211 Geneva 23, Switzerland
\item[\geneva] University of Geneva, CH-1211 Geneva 4, Switzerland
\item[\hefei] Chinese University of Science and Technology, USTC,
      Hefei, Anhui 230 029, China$^{\triangle}$
\item[\lausanne] University of Lausanne, CH-1015 Lausanne, Switzerland
\item[\lyon] Institut de Physique Nucl\'eaire de Lyon, 
     IN2P3-CNRS,Universit\'e Claude Bernard, 
     F-69622 Villeurbanne, France
\item[\madrid] Centro de Investigaciones Energ{\'e}ticas, 
     Medioambientales y Tecnol\'ogicas, CIEMAT, E-28040 Madrid,
     Spain${\flat}$ 
\item[\florida] Florida Institute of Technology, Melbourne, FL 32901, USA
\item[\milan] INFN-Sezione di Milano, I-20133 Milan, Italy
\item[\moscow] Institute of Theoretical and Experimental Physics, ITEP, 
     Moscow, Russia
\item[\naples] INFN-Sezione di Napoli and University of Naples, 
     I-80125 Naples, Italy
\item[\cyprus] Department of Physics, University of Cyprus,
     Nicosia, Cyprus
\item[\nymegen] University of Nijmegen and NIKHEF, 
     NL-6525 ED Nijmegen, The Netherlands
\item[\caltech] California Institute of Technology, Pasadena, CA 91125, USA
\item[\perugia] INFN-Sezione di Perugia and Universit\`a Degli 
     Studi di Perugia, I-06100 Perugia, Italy   
\item[\peters] Nuclear Physics Institute, St. Petersburg, Russia
\item[\cmu] Carnegie Mellon University, Pittsburgh, PA 15213, USA
\item[\potenza] INFN-Sezione di Napoli and University of Potenza, 
     I-85100 Potenza, Italy
\item[\prince] Princeton University, Princeton, NJ 08544, USA
\item[\riverside] University of Californa, Riverside, CA 92521, USA
\item[\rome] INFN-Sezione di Roma and University of Rome, ``La Sapienza",
     I-00185 Rome, Italy
\item[\salerno] University and INFN, Salerno, I-84100 Salerno, Italy
\item[\ucsd] University of California, San Diego, CA 92093, USA
\item[\sofia] Bulgarian Academy of Sciences, Central Lab.~of 
     Mechatronics and Instrumentation, BU-1113 Sofia, Bulgaria
\item[\korea]  The Center for High Energy Physics, 
     Kyungpook National University, 702-701 Taegu, Republic of Korea
\item[\purdue] Purdue University, West Lafayette, IN 47907, USA
\item[\psinst] Paul Scherrer Institut, PSI, CH-5232 Villigen, Switzerland
\item[\zeuthen] DESY, D-15738 Zeuthen, 
     FRG
\item[\eth] Eidgen\"ossische Technische Hochschule, ETH Z\"urich,
     CH-8093 Z\"urich, Switzerland
\item[\hamburg] University of Hamburg, D-22761 Hamburg, FRG
\item[\taiwan] National Central University, Chung-Li, Taiwan, China
\item[\tsinghua] Department of Physics, National Tsing Hua University,
      Taiwan, China
\item[\S]  Supported by the German Bundesministerium 
        f\"ur Bildung, Wissenschaft, Forschung und Technologie
\item[\ddag] Supported by the Hungarian OTKA fund under contract
numbers T019181, F023259 and T024011.
\item[\P] Also supported by the Hungarian OTKA fund under contract
  number T026178.
\item[$\flat$] Supported also by the Comisi\'on Interministerial de Ciencia y 
        Tecnolog{\'\i}a.
\item[$\sharp$] Also supported by CONICET and Universidad Nacional de La Plata,
        CC 67, 1900 La Plata, Argentina.
\item[$\triangle$] Supported by the National Natural Science
  Foundation of China.
\end{list}
}
\vfill



\begin{table}[htbp]
\begin{center}
\begin{tabular}{|c|c|c|c|}                                       \hline
$\sqrt{s}$ (GeV) & Named as & ${\cal L}$ (pb$^{-1}$) & Efficiency (\%) \\ \hline
191.6       & 192 & \phantom{1}28.8 & $64.2 \pm 0.5$ \\
195.5       & 196 & \phantom{1}82.4 & $64.8 \pm 0.2$ \\
199.5       & 200 & \phantom{1}67.5 & $64.7 \pm 0.2$ \\
201.7       & 202 & \phantom{1}35.9 & $64.3 \pm 0.5$ \\
202.5$-$205.5 & 205 & \phantom{1}74.3 & $64.1 \pm 0.2$ \\
205.5$-$209.2 & 207 &           138.1 & $63.6 \pm 0.2$ \\ \hline
\end{tabular}
\end{center}

\icaption{Centre-of-mass energies, luminosities and selection
          efficiencies. Statistical uncertainties from the Monte Carlo sample
          are quoted.
          \label{tab:lumeff}}
\end{table} 

\begin{table}[htbp]
\begin{center}
\begin{tabular}{|c|c|c|c|c|}                             \hline
 ~                & \multicolumn{4}{c|}{Number of events}      \\ \cline{2-5}
$\sqrt{s}$ (GeV)  & \multicolumn{2}{c|}{$2\gamma$} & \multicolumn{2}{c|}{$3\gamma$} \\ \cline{2-5}
~           & Observed & Expected & Observed & Expected \\ \hline
192 & 193 & 207 & \phantom{1}7 & \phantom{1}6  \\
196 & 555 & 575 &           17 &           16  \\
200 & 424 & 453 &           15 &           13  \\
202 & 223 & 236 & \phantom{1}4 & \phantom{1}6  \\
205 & 459 & 464 &           11 &           13  \\
207 & 863 & 845 &           29 &           23  \\ \hline
\end{tabular}
\end{center}

\icaption{Observed and expected number of events with two and three photons.
          \label{tab:number_of_g}}
\end{table}

\begin{table}[htbp]
\begin{center}
\begin{tabular}{|c|c|c|}                                                  \hline
$\sqrt{s}$ (GeV) & $\sigma_{measured}$ (pb) & $\sigma_{expected}$ (pb) \\ \hline
192 & $10.83 \pm 0.74 \pm 0.13$ & $11.5 \pm 0.1$ \\
196 & $10.70 \pm 0.44 \pm 0.10$ & $11.1 \pm 0.1$ \\
200 & $10.05 \pm 0.46 \pm 0.10$ & $10.7 \pm 0.1$ \\
202 & $\phantom{1}9.82 \pm 0.63 \pm 0.13$ & $10.5 \pm 0.1$ \\
205 & $\phantom{1}9.87 \pm 0.45 \pm 0.10$ & $10.0 \pm 0.1$ \\
207 & $10.16 \pm 0.34 \pm 0.10$ & $\phantom{1}9.9 \pm 0.1$ \\ \hline
\end{tabular}
\end{center}

\icaption{Measured and expected cross sections in the angular region 
          $16^{\circ} < \theta_{\gamma} < 164^{\circ}$. For the measured values, the first
          uncertainty is statistical and the second systematic. For the
          expected values the uncertainty due to the missing higher 
          order contributions is estimated to be 1\%.
          \label{tab:xsecs}}
\end{table}

\newpage
\begin{table}
\begin{center}
\rotatebox{90}{
\begin{tabular}{|c|r r r r r r r r|c|}
\hline
 $\cos\theta$ &
\multicolumn{8}{|c|}{Data events/Efficiency{[\%]} ($\sqrt{s}$ in GeV)}
& Radiative correction \\ 
              &   183~~   &   189~~   &   192~~   &   196~~   &   200~~   &   202~~   &   205~~   &   207~~   & factor \\ 
\hline\hline
0.00$-$0.05 & 15/91.7 &  35/87.9 &  5/81.0 &  13/88.4 & 12/87.6 & 10/90.9 & 17/89.1 &  24/88.6 & 0.78 \\
0.05$-$0.10 & 14/89.0 &  21/87.7 &  9/91.7 &  15/85.6 & 14/88.1 &  5/96.7 & 14/85.3 &  28/86.0 & 0.79 \\
0.10$-$0.15 & 10/85.9 &  37/88.1 &  4/82.5 &  10/87.6 &  7/88.8 &  7/86.0 & 11/84.7 &  28/88.7 & 0.80 \\
0.15$-$0.20 &  9/89.4 &  37/87.1 &  7/87.8 &  15/89.6 & 10/85.3 &  5/87.9 & 14/84.3 &  25/88.8 & 0.81 \\
0.20$-$0.25 & 10/90.2 &  46/88.6 &  5/92.1 &  16/88.7 & 15/86.1 &  5/91.4 & 14/86.9 &  15/85.2 & 0.81 \\
0.25$-$0.30 & 18/88.5 &  48/88.4 &  6/80.2 &  20/89.5 & 11/89.7 &  5/91.2 & 12/90.8 &  14/88.7 & 0.82 \\
0.30$-$0.35 & 16/90.7 &  35/86.0 &  0/82.9 &  16/89.0 & 13/86.8 &  8/82.5 &  9/87.4 &  27/89.4 & 0.82 \\
0.35$-$0.40 & 13/88.5 &  45/86.7 &  4/91.6 &  23/89.2 & 16/89.0 &  9/89.6 & 13/92.4 &  24/89.9 & 0.82 \\
0.40$-$0.45 & 13/87.7 &  41/86.0 &  8/77.8 &  19/87.5 & 10/87.2 &  9/92.0 & 17/88.4 &  31/87.9 & 0.83 \\
0.45$-$0.50 & 12/88.5 &  57/88.6 & 10/93.2 &  20/90.3 & 12/89.5 &  7/83.3 & 16/86.8 &  37/89.4 & 0.84 \\
0.50$-$0.55 & 23/88.8 &  74/88.4 &  5/85.2 &  23/87.8 & 14/92.7 &  7/85.5 & 21/88.6 &  47/88.4 & 0.84 \\
0.55$-$0.60 & 17/86.6 &  50/86.6 &  8/84.4 &  20/88.8 & 18/86.1 & 11/84.6 & 27/84.4 &  41/87.7 & 0.85 \\
0.60$-$0.65 & 31/82.5 &  73/82.9 & 10/82.6 &  31/84.1 & 26/85.1 & 15/82.9 & 24/86.4 &  47/82.1 & 0.86 \\
0.65$-$0.70 & 21/77.7 &  66/77.9 &  9/76.8 &  29/77.5 & 32/78.3 & 15/76.7 & 28/76.3 &  61/75.2 & 0.87 \\
0.70$-$0.75 &  8/17.0 &  27/16.3 &  2/15.4 &  11/17.3 &  7/17.8 &  6/16.0 &  9/16.5 &  10/16.7 & 0.87 \\
0.75$-$0.80 &  5/14.3 &  20/13.5 &  2/11.6 &  11/12.3 & 10/14.7 &  3/14.9 &  5/13.2 &  20/12.6 & 0.88 \\
0.80$-$0.85 & 38/53.5 & 103/52.5 & 19/55.8 &  41/53.2 & 27/49.7 & 20/47.1 & 40/52.1 &  61/50.4 & 0.89 \\
0.85$-$0.90 & 78/79.8 & 223/80.7 & 26/73.6 &  92/74.9 & 74/74.3 & 33/74.9 & 72/76.3 & 137/76.7 & 0.91 \\
0.90$-$0.95 & 73/66.8 & 258/66.6 & 45/65.6 & 114/66.0 & 83/66.0 & 36/67.4 & 83/63.9 & 154/63.7 & 0.95 \\
0.95$-$0.96 & 35/69.1 &  78/67.2 & 16/67.4 &  33/66.7 & 28/66.3 & 11/66.1 & 24/63.7 &  61/62.9 & 1.00 \\
\hline
\end{tabular} }
\end{center}
\icaption{Number of events, efficiency and radiative correction factor 
          applied to the data as a function of $\rts$ and of the event polar 
          angle, $\cos\theta$. The
          values at $\sqrt{s}=183$ and $189 \GeV$~{\protect \cite{gg189}} are 
          also listed. The uncertainty
          on the radiative correction factor ranges from 5\% (first 
          $\cos\theta$ bin) to 1\% (last $\cos\theta$ bin) and is due to the 
          finite Monte Carlo statistics. \label{tab:lep}}
\end{table}

\newpage
\newpage
\newpage
\newpage
\newpage

\end{document}